# Post-processing techniques of 4D flow MRI: velocity and wall shear stress

Qi Gao[1], Xingli Liu[2], Hongping Wang[3], Fei Li[4,*], Peng Wu[5,*], Zhaozhuo Niu[6,*], Mansu Jin[3], RunJie Wei[3]

1: School of Aeronautics and Astronautics, Zhejiang University, Hangzhou, China;
2: Hangzhou Shengshi Technology Co., Ltd., Hangzhou, China;
3: Microvec. Inc., Beijing, China;
4: Department of Structural Heart Disease, Fuwai Hospital; State Key Laboratory of Cardiovascular Disease; Chinese Academy of Medical Sciences and Peking Union Medical College; Beijing, China
5: Artificial Organ Technology Lab, Bio-manufacturing Research Centre, School of Mechanical and Electric, Engineering, Soochow University, Suzhou, China;
6: Cardiac Surgery, Qingdao Municipal Hospital, Qingdao, China

Correspondence Authors: Fei Li, leestein@126.com; Peng Wu, pwu@suda.edu.cn; rudolfniu@163.com

**Purpose:** As the original velocity field obtained from four-dimensional (4D) flow magnetic resonance imaging (MRI) contains considerable amount of noises and errors, the available Divergence-free smoothing (DFS) method can be used to process the 4D flow MRI data for reducing noises, eliminating errors, fixing missing data and eventually providing the smoothed flow field. However, the traditional DFS does not have the ability to deal with the flow in the near wall region of vessel, especially for satisfying the no-slip boundary condition. In this study, therefore, an improved DFS method with specific near wall treatment is introduced for processing with 4D flow MRI inner flow with curved wall boundary as the blood flows. On the other hand, due to the coarse resolution of 4D flow MRI, velocity gradients in the near wall region are normally underestimated. As a result, a special wall function is required for accurately computing wall shear stress (WSS).

**Methods:** Firstly, optimized objective function was built according to the constraints that boundary condition of wall was no-slip and flow field satisfied divergence-free requirement. Then objective function was minimized to establish linear equation group with respect to velocity and smoothing parameters. Secondly, function of minimized generalized cross validation (GCV) was introduced to optimize the smoothing parameters. Subsequently, smoothing parameters was brought back to velocity equation to achieve velocity distribution which satisfies the divergence-free restraint and smoothness in velocity field. Finally, according to the near-wall local coordinate system, near wall velocity distribution was fitted by binomial using

Musker wall function to calculate WSS, in which the Musker wall function was derived from eddy viscosity model and it was capable of resolving near-wall velocity gradient accurately.

**Results**：During systole, there was a high-speed flow region with maximum velocity at the lateral side of ascending aorta and helical flow at the inner curvature for both data of 4D flow MRI and Computational fluid dynamics (CFD). The reason is the formation of low-pressure zone and the emergence of flow separation at the ascending aorta near the inner side, which was mainly because the patient suffered from aortic regurgitation. Compliance in vivo leads to the apparent difference between 4D flow MRI and CFD in velocity distribution in aortic arch. For the flow in supra-aortic vessels, velocity magnitude from CFD was higher than that from 4D flow MRI. It meant that the accuracy of velocity profile was still limited by the insufficient resolution in 4D flow MRI, which led to the underestimated WSS at supra-aortic trunks. At the same time, it also suggested that velocity field indeed was overestimated in numerical simulation in supra-aortic vessels compared with 4D flow MRI results, which also resulted in the overestimated WSS at supra-aortic arteries. The potential reason could be that the vascular compliance and unsteady effect of blood flow were not concerned in the numerical simulation. Except that, the WSS calculated by 4D flow MRI was significantly optimized with the improved DFS and proposed wall treatment.

**Conclusions**：In summary, the improved DFS can greatly reduce the divergence error of the entire flow field, so that flow field satisfies the no-slip condition on the wall and divergence-free requirement. Velocity distribution is restored well and aortic WSS is correctly calculated to a large extent based on 4D Flow MRI by the comparison with CFD. This is helpful for judging the location and the possible development of lesions of diseases.

**KEYWORDS**

4D flow MRI, Computational fluid dynamics, Wall shear stress, Divergence-free smoothing

# INTRODUCTION

**4D flow MRI**

In the 1980s, the concept of phase contrast (PC) magnetic resonance imaging (MRI) was proposed, in addition to fluoroscopic imaging, to measure the velocity field of flow. A new term was brought up when MRI was used in measurement in fluid dynamic velocity field, that is Magnetic resonance velocimetry (MRV). This technique was widely used in applications where optical measurements are difficult to achieve, such as non-transparent in vivo flow, porous media seepage, and two-phase flow. In the meantime, MRI measurements are also important in the medical field. It has been widely accepted by clinicians due to its unique advantage in the visualization of blood flow and quantitative assessment of hemodynamics in the heart, aorta and other large blood vessels[1,2].

MRI module uses a special gradient technique to compute the average velocity of the MR image data, which can be utilized on common medical MRI scanners[3]. This time-resolved flow measurement can be traced back to small blood vessels, such as MR angiography measurements in coronary arteries. However, measurement data is severely distorted by reason of respiration action, resulting in flow blur and artifacts[4]. To address this problem, Hennig et al. (1988) presented some image sequences changed over time by echocardiographic gating. Those images were capable of resolving a series of velocity-encoded three-directional (3D) velocity fields, which enabled time-resolved visualization of blood flow[5,6]. Due to the versatility of this technology, the widespread application of MRI scanners, as well as the non-invasive property, the elimination of light sources and tracer particles, MRI velocity has been extensively used in the investigation of blood flow in the medical field[7,8].

In recent years, 4D flow MRI has been getting popular in vivo research and clinical medicine as a new diagnostic tool, referring to 3D time-resolved PC-MRI with three-directional flow encoding[9]. In addition to providing morphological information, it also permits the acquisition of functional information and further obtains the hemodynamic index such as velocity[10], energy loss (EL)[11], pressure differences[12,13] as well as wall share stress (WSS)[14] within a 3D data acquisition vascular region of interest. It has been proved that 4D flow MRI shows an association between blood flow and cardiovascular and cerebrovascular diseases[15,16].

## Application of MRI technique for large vessels

Increasingly, 4D flow MRI is extensively studied in the clinic. In addition to the most prominent aortic lesions[17,18], it also investigates ventricles[19–21], atriums[22–24], heart valves[25,26], pulmonary arteries and veins[27–29], carotid arteries[30–32], intracranial arteries and veins[33–35], hepatic arteries and portal veins[36,37], peripheral blood vessels[38] and renal arteries[39] and other organs.

Common aortic diseases include aortic aneurysm, aortic dissection, atherosclerosis, and aortic inflammation etc[40]. The formation of these diseases may be related to the blood flow pattern inside the aorta, and the occurrence of pathology will further change the pattern of blood flow, thereby aggravating the lesion or triggering new ones. For instance, by comparing the aortic blood flow in healthy people and in patients with ascending aortic aneurysm, it is found that helical flow was larger, while the retrograde flow occurs earlier and lasts longer in the patients' aorta. Meantime, the average velocity between the ascending aorta and the transverse aorta is much higher in aneurysm patients than that in volunteers[16]. The aortic blood flow patterns are also strongly related to the lesions of the aorta connecting blood vessels and organs. For example, by observation of the blood flow patterns in the aorta of patients with coronary artery disease (CAD) by means of MRV technology, it is discovered that the trend of atherosclerosis increases with age. Moreover, the blood flow in the ascending aorta is irregular in both normal elderly and all patients, which in turn leads to a decrease in blood flow into the coronary arteries[41]. In the past few years, wall shear stress (WSS) obtained by 4D flow MRI has become a new diagnostic indicator for disease. A large amount of data indicates that WSS, like other hemodynamic indicators, plays an important role in the development and formation of vascular diseases[11,42,43]. WSS is regarded as the most likely reason to be responsible for the dilatation of the aorta or the formation of aortic aneurysm[44,45]. For instances, local low or high WSS may respectively promote or prevent atherosclerotic lesions in the aortic wall. To better depict the characteristics of the saccular aneurysms and fusiform aneurysms, a recent study was performed by Natsume et al. (2017) who evaluated the relation between the geometry of aortic arch aneurysms and WSS, and vortex flow utilizing 4D flow MRI in 100 patients. This study found that vortex flow was always present in the aneurysms, resulting in low WSS. The results show that fusiform aneurysms elongate as they dilate, and WSS is lower as the diameter is larger. Saccular aneurysms dilate without proportionate elongation, in which those attached inner curvatures have low WSS regardless of diameter and may behave malignantly[46].

4D flow MRI can measure the velocity field in all directions in space and has been applied in clinic, but there are still some shortcomings in this technique. Since the data is obtained through experimental measurement, it contains experimental error or noise which can affect the measurement accuracy. At present, there is no effective post-processing technology for 4D flow MRI to reduce the noise and suppress the error. On the other hand, it is still lacking mature techniques to extract clinically valuable hemodynamic physiological indicators via velocity field measured by 4D flow MRI. For example, the restoration of the near-wall velocity field, especially the accurate prediction of velocity gradient, has a straightforward impact on the calculation of the important physiological index of WSS. Therefore, it is significant to optimize the 4D flow MRI measurement data from velocity field post-processing and near-wall velocity distribution modeling.

**Review of flow field data post-processing techniques**

In the field of experimental fluid mechanics, velocity field can be obtained by particle image velocimetry (PIV) technique[47]. With the development of tomography PIV[48,49], 3D3C (three-dimensional three-component) and even 4D3C (time-resolved 3D3C) methods have become the trend of flow field measurement as it can acquire more physical information. Post-processing technique of velocity field is one of the most important research interests on PIV measurement. It generally consists of three steps: data verification, data interpolation, and data smoothing[50]. Data verification is usually used to identify velocity outliers in the data. The normalized median test proposed by Westerweel and Scarnao (2005) [51] as one of the data verification methods, was substantially used in PIV, which can adaptively identify the flow field error vector with high accuracy and robustness. Error velocity vectors are usually replaced by linear, spline or Kriging interpolation methods[52]. Data smoothing can effectively eliminate random errors, which is usually realized by simple digital filtering. In order to improve the efficiency of data post-processing, Garcia (2011) proposed DCT-PLS method combined with minimum penalty least squares (PLS) optimization and discrete cosine transform (DCT), which can fully automate data verification, interpolation and smoothing[50].

Velocity field obtained by 4D flow MRI measurement is similar to the PIV velocity field, which should satisfy the physical law (or governing equations), such as Naiver-Stokes equation (NS equation). 4D flow MRI data can be processed mathematically well by the above-described optimization method similar to processing PIV data, but the obtained result does not actually satisfy the NS equation. Therefore, mass conservation as physical constraint, that is divergence-free condition, can be used to correct 4D flow MRI data. For instance, Song et al. (1993) projected the MRI velocity field into a divergence-free space by finite difference method to achieve the goal of improving signal-to-noise (S/N)[53]. Busch et al. (2013) and Ong et al. (2013) implemented fast and efficient divergence-free correction of the MRI velocity field by divergence-free radial basis functions and diverging-free wavelets, respectively[54,55]. In the aspect of PIV, the divergence-free condition is also utilized to correct the measured velocity field. For example, De Silva et al. (2013) proposed the divergence correction scheme (DCS) method, which solves the nonlinear equations to enable the measurement flow field to satisfy the divergence-free condition, and to ensure the minimum deviation between the revised flow field and the measured flow field[56]. For the sake of smoothing flow field simultaneously, Wang et al. (2016) proposed the divergence-free smoothing (DFS) method in combination with DCT-PLS[50,57]. Although 4D flow MRI and PIV are fundamentally different in measurement principle, both data structures are very similar, which means that 4D flow MRI flow field and PIV flow field are both suffering from the same problem that the measured flow field contains strong noise and low temporal and spatial resolution[58], which affects further visualization of flow structure and flow field quantitative analysis. Therefore, by drawing lessons from the experience of PIV post-processing, 4D flow MRI flow field can be optimized through adopting basic physical constraints during the post-processing procedure.

The aim of MRI flow field post-processing mainly is to improve the ability of data visualization and the calculation accuracy of physiological index. Currently, WSS is one of most common physiological indicators for MRI data analysis, which reflects the shear stress of blood flow on the vessel wall and is positively correlated with the velocity gradient in the near wall region. WSS can be estimated from temporal and spatial information on near wall velocity direction and magnitude. In the early years, based on blood vessel 3D cine PC-MRI data, a method for calculation of WSS was presented on 2D cross-section sliced of the vessel model, in which B-splines interpolation method was used for blood flow velocity field fitting[59]. According to the inner normal direction of the wall point, numerical differentiation and a linear least squares method were utilized, respectively, to evaluate the spatial derivative of tangential velocity on the wall, and then 3D volumetric WSS was computed[35,42]. Compared with 2D slices WSS, the practicality of 3D volumetric WSS is significantly improved, meanwhile, the robustness of boundary velocity fitting is also enhanced. However, some issues of inaccurate vascular segmentation, low resolution, signal-to-noise ratio, spatial filtering, and even the choice of encoding velocities, have a severe impact on the calculation accuracy of WSS. Therefore, relying on the vascular model and velocity vector distribution, Potters et al. (2015) acquired the velocity distribution near the vessel wall using the cubic spline curve of natural adjacent point interpolation[43]. Based on the body pixel grayscale and the collected blood flow velocity, Riminarsih et al. (2016) restored the velocity of each pixel position inside the aorta using blood flow calculation formula which was proposed by Xavier (2007)[60]. Using the least squares method, velocity profile near the wall were fitted to the paraboloid by applying blood velocity within the 80% to 95% radius of the aorta[61]. Potters et al. (2015) and Riminarsih et al. (2016) built relatively accurate velocity profiles and obtained the WSS, but the physical methods they introduced both ignored the effect of fluid viscosity near the wall. Considering the profiles of blood velocity in the log-linear region, WSS was calculated from the near-wall velocity which was fitted by the eddy viscosity model[62] in this paper.

In the current work, the objective function of velocity optimization is defined according to the constraints of no-slip wall and divergence-free practice. Then the establishment of linear equation system with respect to velocity and smoothing parameters coming from minimizing the objective function is aimed to acquire the aortic velocity field which satisfies the divergence-free constraint and smoothness. Based on constructed local coordinate system, the near-wall velocity profile is fitted by using Musker wall function and then the WSS is calculated. Thus, the correction of flow results and WSS coming from 4D flow MRI can be evaluated by comparison of the calculation results computed from CFD.

## METHODS

A detailed investigation from geometry reconstruction based on MRI data of blood flow visualization for WSS comparison between 4D Flow MRI and CFD was carried out to evaluate feasibility of improved post-processing techniques. Written informed consent was obtained from the participant and this study was approved by the local ethics committee.

**Patients and MRI data acquisition**

A male patient, aged 56 years old, with ascending aortic aneurysm complicated with aortic regurgitation and aortic root aneurysm was selected for this study. No other conspicuous cardiovascular disease was found in previous diagnoses in hospital. The rest of the required baseline characteristics can be collected from discharge papers. MRI was performed on the patient using a 3.0 Tesla GE MR scanner and Cardiac-gated 4D Flow MRI sequences were obtained with respiratory synchronization method. Imaging parameters were as follows: TR/TE = 4.8/2.3 ms, Flip angle = 10 degree, FOV = 320×263 mm, slice thickness = 2.2 mm, View per cardiac phase = 25, Temporal resolution = 38.4 ms, Image matrix size = 256×256×80, Voxel size = 2.1×2.4×2.2 mm$^3$. With the data, 4D Flow MRI can conduct the phase encoding on three mutually perpendicular dimensions simultaneously to collect the blood flow data.

**Morphologic reconstruction**

4D flow MRI velocity field is consisted of a total number of 80×3 images, which are velocity component contours in three directions. Two neighborhood methods, named neighborhood variance method and neighborhood sign determination method, are respectively used to denoise all images, which can remove most of the noise and allow the appearance of crude outline of the aorta. The formulas of neighborhood variance method (Eq. 1) and neighborhood sign determination method (Eq. 2) are given as follows:

$$I_S(i,j) = \sqrt{\frac{1}{mno} \sum_{(x,y) \subset S} (f_S(x,y) - \overline{f}_S)^2}, \quad (1)$$

and

$$PN_S(i,j) = |length_S(P) - length_S(N)|^2. \quad (2)$$

Where, $S$ is the region of neighborhood, $m$ and $n$ are dimensions of $S$, $(x,y)$ is the local coordinates of $S$, $f_S(x,y)$ is the intensity of point $(x,y)$, $\overline{f}_S$ is the average intensity in $S$, $I_S(i,j)$ is the variance in $S$ centered at point $(i,j)$, $(i,j)$ is the global coordinates of the raw image, $length_S(P)$ is the number of pixels with positive intensity in $S$, $length_S(N)$ is the number of pixels with negative intensity in $S$.

All residual noise is eliminated using median filter through the combination of two neighborhood denoised images[63]. Then, automatic threshold segmentation of images is proceeded using Otsu's method to produce ultimate images which can be used to extract aorta model[64]. A 3D model with the form of points cloud is preliminarily generated by merging all of denoised images with the pattern of original slices of 4D flow MRI.

Finally, 3D aortic morphological geometry is obtained by using the Poisson surface reconstruction which transforms the points cloud into surface[65], as shown in Fig. 1.

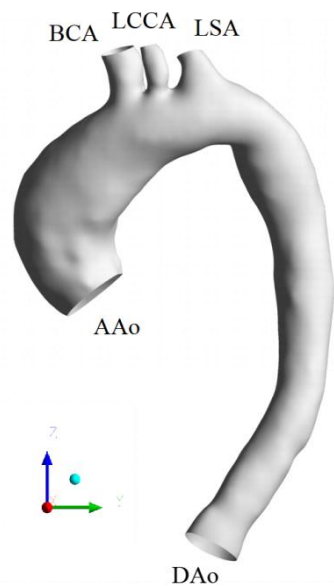

Fig. 1 Geometry construction

**New model of near-wall velocity for WSS calculation**

Flow in blood vessels measured by 4D flow MRI is incompressible flow. The velocity field of blood flow should satisfy the conservation of mass, namely the divergence-free condition. Therefore, DFS method can be applied to the 4D flow MRI for error reduction. Under the hypothesis that blood vessel is inelastic, velocity of 4D flow MRI at the vessel wall is zero according to the no-slip boundary condition. However, the original DFS method does not provide a near wall treatment satisfying boundary conditions[57], and the greatest difficulty in applying DFS to 4D flow MRI is how to deal with no-slip wall condition. This section will detail the process of establishing DFS equations with wall boundary condition through a blood vessel model.

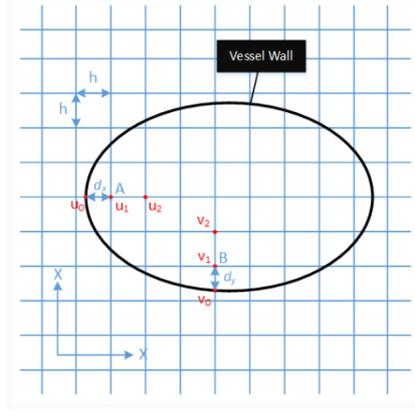

Fig. 2 2D schematic of velocity grid and vessel wall

Fig. 2 shows the 2D diagram of velocity grid and vessel wall (solid black curve). The velocity of 4D flow MRI locates at the inner grid. When applying DFS to 4D flow MRI, the most important thing is to establish the difference equation in the near wall region. Take the two points A and B in the figure as an example and they are the nearest points along the normal directions of the wall with the spacing to the wall as $dx\,(dx>0)$ and $dy\,(dy>0)$, respectively. Both of them are smaller than the grid spacing $h$, which means the location of the wall has a sub-grid precision. Taylor expansion of velocities at point A is

$$u_0 = u_1 - \left(\frac{\partial u}{\partial x}\right)_0 dx + \frac{1}{2}\left(\frac{\partial^2 u}{\partial x^2}\right)_0 dx^2 + \mathrm{O}(dx^3), \tag{3}$$

$$u_2 = u_1 + \left(\frac{\partial u}{\partial x}\right)_0 h + \frac{1}{2}\left(\frac{\partial^2 u}{\partial x^2}\right)_0 h^2 + \mathrm{O}(h^3). \tag{4}$$

Let $\theta = dx/h$, the derivatives in Eq. (3) and (4) become

$$\left(\frac{\partial u}{\partial x}\right)_0 = \frac{1}{h}\left(-\frac{1}{\theta^2+\theta}u_0 + \frac{1-\theta}{\theta}u_1 + \frac{\theta}{1+\theta}u_2\right), \tag{5}$$

$$\left(\frac{\partial^2 u}{\partial x^2}\right)_0 = \frac{2}{h^2}\left(\frac{1}{\theta^2+\theta}u_0 - \frac{1}{\theta}u_1 + \frac{1}{1+\theta}u_2\right). \tag{6}$$

Similar expansions of velocities at point B could be achieved. For the grid points in the near wall region, the difference equations need to be re-established according to equations (5) and (6), while the central difference scheme is applied at the inner points far from the wall boundary.

In the DFS method, the divergence of smoothed velocity field needs to be zero. At the same time, the difference from the original velocity field should be as small as possible. Consequently, the objective function of the optimization process is[57,66]:

$$J(\mathbf{U}) = (\mathbf{U} - \mathbf{U}_m)^T (\mathbf{U} - \mathbf{U}_m) + sR(\mathbf{U}) \quad , \quad \text{subject to} \quad \nabla \cdot \mathbf{U} = 0 \tag{7}$$

Where, $\mathbf{U}$ is the column vector $[\mathbf{u}, \mathbf{v}, \mathbf{w}]^T$ consisting of three velocity components, $(\mathbf{U} - \mathbf{U}_m)^T (\mathbf{U} - \mathbf{U}_m)$ represents residual sum-of-square (RSS) between the optimization velocity and the original measurement velocity, $R(\mathbf{U})$ reflects the smoothness of the velocity field, smoothing parameter $s$ is a positive value, and the larger $s$ means the smoother velocity field. In this paper, the second derivative of velocity is used to characterize the smoothness of the flow field, namely:

$$R(\mathbf{U}) = \|\mathbf{D}\mathbf{U}\|^2 = \mathbf{U}^T \mathbf{D}^T \mathbf{D} \mathbf{U}. \tag{8}$$

Where, $\mathbf{D} = \frac{\partial^2}{\partial x^2} + \frac{\partial^2}{\partial y^2} + \frac{\partial^2}{\partial z^2}$ is discrete second-order derivative operator, $\nabla \cdot \mathbf{U}$ is the divergence of velocity, which can be reformatted with discrete divergence operator $\mathbf{A}\mathbf{U}$. Note that the operator $\mathbf{A}$ and $\mathbf{D}$ in the near wall region need to be discrete in terms of equations (5) and (6), while the central difference scheme will be used in the inner region.

According to the Lagrange multiplier method, the original objective function $J(\mathbf{U})$ can be transformed into a new format $L(\mathbf{U})$:

$$L(\mathbf{U}, \lambda) = (\mathbf{U} - \mathbf{U}_m)^T (\mathbf{U} - \mathbf{U}_m) + sR(\mathbf{U}) + 2\lambda^T \nabla \cdot \mathbf{U}. \tag{9}$$

Where $\lambda$ is the Lagrangian multiplier. The first derivative of $\mathbf{U}$ and $\lambda$ is set to be zero. Therefore, the following linear equations can be obtained by minimizing the objective function,

$$\begin{cases} (\mathbf{I}+s\mathbf{D}^\mathrm{T}\mathbf{D})\mathbf{U}+\mathbf{A}^\mathrm{T}\lambda = \mathbf{U}_m \\ \mathbf{A}\mathbf{U}=0 \end{cases}. \tag{10}$$

Where $\mathbf{I}$ is a unit matrix, and further arrangement to the linear equations show as,

$$\begin{bmatrix} \mathbf{I}+s\mathbf{D}^\mathrm{T}\mathbf{D} & \mathbf{A}^\mathrm{T} \\ \mathbf{A} & 0 \end{bmatrix} \begin{bmatrix} \mathbf{U} \\ \lambda \end{bmatrix} = \begin{bmatrix} \mathbf{U}_m \\ 0 \end{bmatrix}. \tag{11}$$

When the smoothing parameter $s$ is given, the optimized velocity field $\mathbf{U}$ can be obtained by solving the linear equations. The equation (11) can be solved by iterative method, in which the coefficient matrix is a sparse matrix.

Since the noise level of the flow field in 4D flow MRI cannot be estimated in advance, the smoothing parameter $s$ in equation (11) needs to be automatically determined based on the flow field data. The choice of s is optimized by minimizing the generalized cross validation (GCV) function following the method by Garcia (2010)[66]. The GCV function is defined as follows:

$$GCV(s) = \frac{(\mathbf{U}-\mathbf{U}_m)^\mathrm{T}(\mathbf{U}-\mathbf{U}_m)/3n}{\left\{1-Tr\left[(\mathbf{I}+s\mathbf{D}^\mathrm{T}\mathbf{D})^{-1}\right]/3n\right\}^2}, \tag{12}$$

where $n$ is the number of unknowns, $Tr$ represents the trace of matrix. According to the character of matrix, $Tr\left[(\mathbf{I}+s\mathbf{D}^\mathrm{T}\mathbf{D})^{-1}\right]$ can be simplified to the following form:

$$Tr\left[(\mathbf{I}+s\mathbf{D}^\mathrm{T}\mathbf{D})^{-1}\right] = \sum_{i=1}^{3n} \frac{1}{1+s\lambda_i}, \tag{13}$$

where $\lambda_i$ is the eigenvalue of the sparse matrix $\mathbf{D}^\mathrm{T}\mathbf{D}$. Due to the large amount of data in a single velocity field of 4D flow MRI, the efficiency of direct equation solving will be very low. In the current work, an approximate solution method is adopted. Since the eigenvalues of $\mathbf{D}^\mathrm{T}\mathbf{D}$ obey the exponential decay law approximately, the eigenvalues are fitted by exponential functions for the purpose of quick solution. Firstly, the front 200 maximum eigenvalues are solved accurately. Secondly, the eigenvalue distribution is fitted by exponential function. Finally, each eigenvalue is calculated approximately according to the fitting result. After finding the

$s$ which makes the $GCV(s)$ smallest, the velocity field that satisfies both the non-discrete constraint and the smoothness can be acquired by bringing $s$ back to the formula (11).

**WSS estimation**

The vessel geometry model is discretized into triangular meshes on the surface. The WSS can be evaluated from the velocity field at each mesh gird. Before calculating WSS, we construct a local coordinate system at each mesh grid as shown in Figure 3.

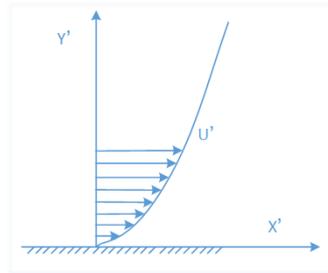

Fig. 3 The local coordinate system

In the figure, $X'$ is the direction of flow velocity, $Y'$ is the normal direction of wall, and the origin is located at the mesh grid. The procedure of building the local coordinate system is as follows:

1) The spline interpolation is applied to the velocity field along the normal direction. In this study, the interpolation spacing between two adjacent points is equal to the velocity field grid spacing. The first interpolation point located on the wall and its velocity is set to zero.

2) The $Z'$ direction of the local coordinate system is computed, which should be perpendicular to the wall normal and velocity vector.

3) Flow direction of $X'$ is obtained by cross-multiplying the $Y'$ direction and the $Z'$ direction.

4) The interpolated velocity is projected to the flow direction $X'$ to get the velocity distribution of $U'$ in the local coordinate system. Note that velocity on the wall is zero.

5) At last, the profile of velocity $U'$ is fitted using Musker model[42,43,62,67] as shown in Eq. (14):

$$U' = u_\tau \int_0^{Y^+} \frac{\frac{(Y^+)^2}{k} + \frac{1}{s}}{(Y^+)^3 + \frac{(Y^+)^2}{k} + \frac{1}{s}} dY^+ \quad , Y^+ = \rho u_\tau Y'/\mu \quad (14)$$

This formula is based on eddy viscosity model, in which the parameters, $u_\tau$、$\rho$ and $\mu$ are the wall friction velocity, the blood density, and the dynamic viscosity coefficient of the blood, respectively, and the constants $k$ and $s$ are 0.41 and 0.001093, respectively. The only unknown parameter in equation (14) is the wall friction velocity $u_\tau$, and its optimal value is obtained by curve fitting. Then $u_\tau$ is brought into the formula (15) to calculate the WSS and the direction of the WSS is consistent with the direction of $X'$.

$$\tau = \rho u_\tau^2 \quad (15)$$

**CFD method**

In this study, numerical simulation is performed to provide references and comparison for the flow fields obtained from 4D flow MRI. The software of ANSYS ICEM-CFD (ANSYS Inc. Canonsburg, PA, USA) is applied to mesh the aortic arch model. Tetrahedral elements mesh is generated because of the complexity of aorta geometry. Prism cells are applied to capture the flow boundary layer, in which the thickness of first layer is 0.03mm. All mesh quality is far above 0.2 in terms of orthogonality, which is usually considered as an acceptable threshold value for CFD calculation. Three different mesh sizes are generated with 0.8 million cells, 1.6 million cells and 3.0 million cells to test the mesh-independency. The results suggest that in addition to ascending and descending aorta, there is a big difference in the prediction of velocity in core regions and WSS between 0.8 million cells mesh and other two meshes (results not shown). Considering the computational cost, the 1.6 million cells mesh is therefore chosen for the following calculations.

Regarding the boundary conditions for simulation, the velocity distributions are extracted at the entrance of aorta at the systolic peak (fifth timestep of cardiac cycle) according to the 4D flow MRI flow field data at 30 moments in one cardiac cycle, which is imposed on the entrance of the computational model and interpolated to the mesh points as the actual inlet velocity boundary condition. The outlets include brachiocephalic artery (BCA), left common carotid artery (LCCA), left subclavian artery (LCA) and descending aorta (DAo). An extension with ten times length of the vessel diameter is added at the aorta outlet boundary to minimize numerical problems of convergence and reverse flow, which is normally caused from flow separation and the loss of pressure of distal end. Meanwhile, Windkessel model is introduced to simulate the whole aortic circulation (see Fig. 4). The capacitance C is considered to be zero because velocity at inlet boundary is a constant profile, and average $R_p$ and $R_d$ are used for various four aortic outlets[68,69]. It is noticed that the aortic wall is inelastic and blood flow cannot pass.

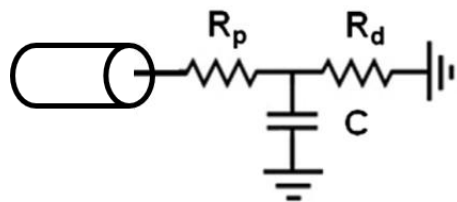

Fig. 4 Windkessel model coupled on different four aortic outlets

The blood flow is numerically simulated by solving the Navier-Stokes equations in ANSYS Fluent 19.2 (ANSYS Inc. Canonsburg, PA, USA). The blood is assumed to be non-Newtonian fluid, in which the density is $\rho = 1060 \, kg/m^3$ and the viscosity is $\mu = 0.0035 \, Pa \cdot s$. The 'SIMPLE' method is utilized as the coupling of pressure and velocity method, and the second order implicit method is used to discretize the transient formulation. The $k - \varepsilon$ RNG model is applied as turbulent model.

## RESULTS

**Post-processing of 4D Flow MRI with improved DFS**

In order to verify the correctness and accuracy of DFS method with wall boundary conditions, three 4D flow MRI velocity fields processed with different approaches are investigated. They are original 4D flow MRI velocity field, 4D flow MRI velocity field smoothed with traditional DFS method and improved DFS method, respectively, as shown in Fig. 5. It is needed to point out that because traditional DFS method cannot deal with the wall boundary conditions, the obtained results are not credible on the wall. It is found that original 4D flow MRI velocity field demonstrates the pattern of blood flow in the core region of aorta. However, the local suspected velocity vector can be seen on the lateral sections (Fig. 5(a)). Fig. 5(b) shows the 4D flow MRI velocity field smoothed by traditional DFS method, in which the noise is reduced evidently, but the near-wall velocity is still untreated and the lack of velocity vector in the core region of descending aorta is also notable. As can be seen from Fig. 5(c), 4D flow MRI velocity field smoothed by the improved DFS can not only correctly reserve the original velocity in the mainstream region, but also improve the velocity field near the wall with better resolved velocity gradient. Table 1 gives the statistical results of mean value and maximum value of the divergence under three different circumstances. The value of divergence reaches the highest in original 4D flow MRI velocity field, and it decreases when traditional DFS method is used to optimize velocity field. The improved DFS method manifests the minimum divergence that can be reached among three methods. The difference can also be seen from the distribution of divergence in three vertical sections from Fig. 5(e, f, g) that traditional DFS cannot reduce the divergence error especially on near the wall.

Table 1. Divergence of three various flow filed

|  | Original data | Traditional DFS | Improved DFS |
| --- | --- | --- | --- |
| Average | 0.1164 | 0.0623 | 0.0074 |
| Maximum | 0.2163 | 0.1813 | 0.0087 |

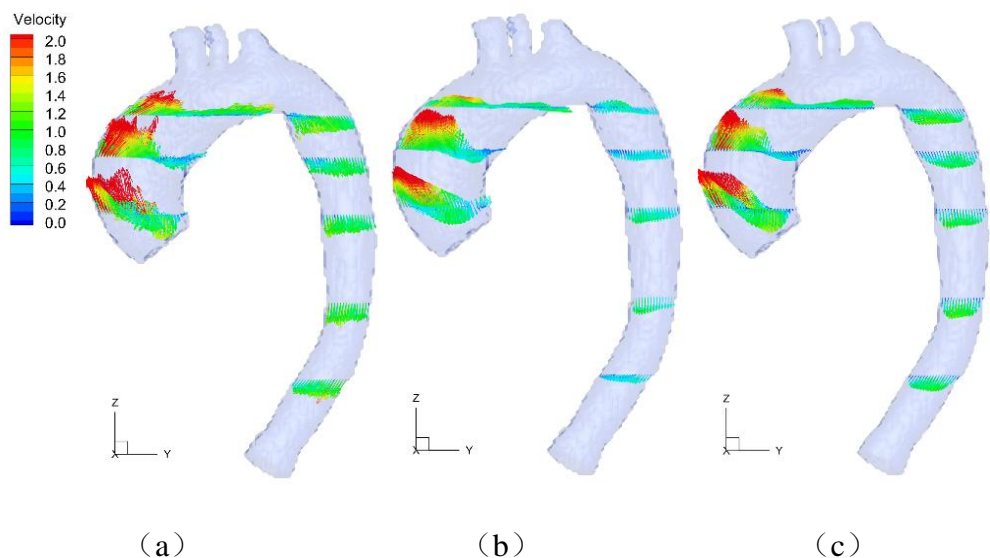

（a） （b） （c）

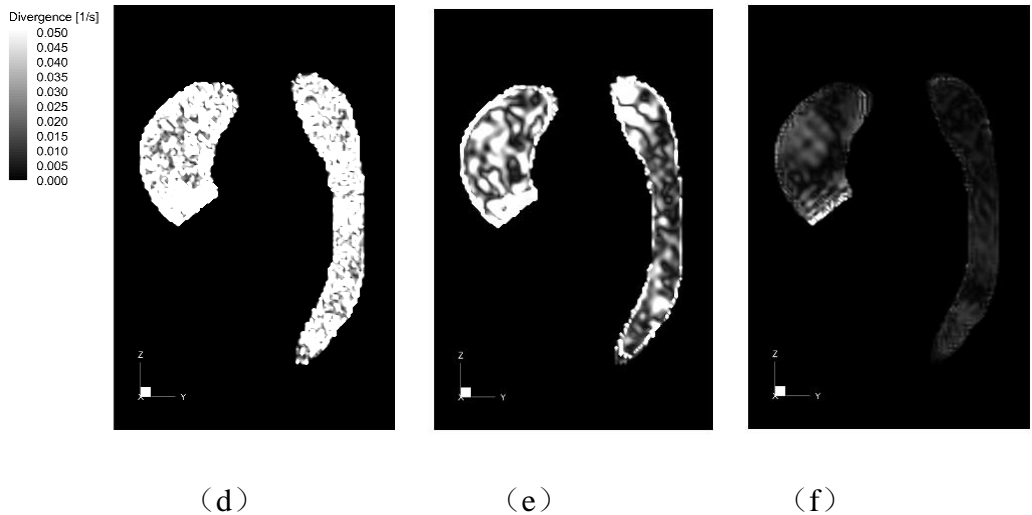

(d) (e) (f)

Fig. 5 The comparison of local velocity profile and divergence distribution on three circumstances: (a, d) original 4D flow MRI, (b, e) 4D flow MRI optimized by traditional DFS method, (c, f) 4D flow MRI optimized by improved DFS method

**Comparison of velocity with CFD simulation**

Fig. 6 shows the streamlines of blood flow, velocity vector distribution and velocity contour maps in three characteristic lateral sections at the peak of systole from optimized 4D flow MRI velocity field and CFD computing result. On the medial side of the ascending aorta, both 4D flow MRI and CFD show back-flow and swirling flow. A high-speed flow region is formed at the greater curvature side of ascending aorta, where blood flow reaches the maximum. There is spiral flow in CFD and subtle flow separation in 4D flow MRI in the anterior segment of the descending aorta (close to Plane 3). At the distal of descending aorta, velocity magnitude in 4D flow MRI is slightly less than that in CFD. Plane 1 and Plane 3 display a great consistency and similarity in velocity magnitude and distribution of 4D flow MRI and CFD. For plane 2, the velocity distribution in 4D flow MRI is bilaterally symmetric which is quite different from CFD, and velocity magnitude at LCCA is much smaller in 4D flow MRI than that in CFD.

Table 2 shows the Pearson correlation coefficient and the average error of velocity in three characteristic planes extracted from both 4D flow MRI and CFD. The correlation profile between three different velocity components are shown in Fig. 7. Except for Plane 2, the correlations of velocity components are sufficiently high in other two planes.

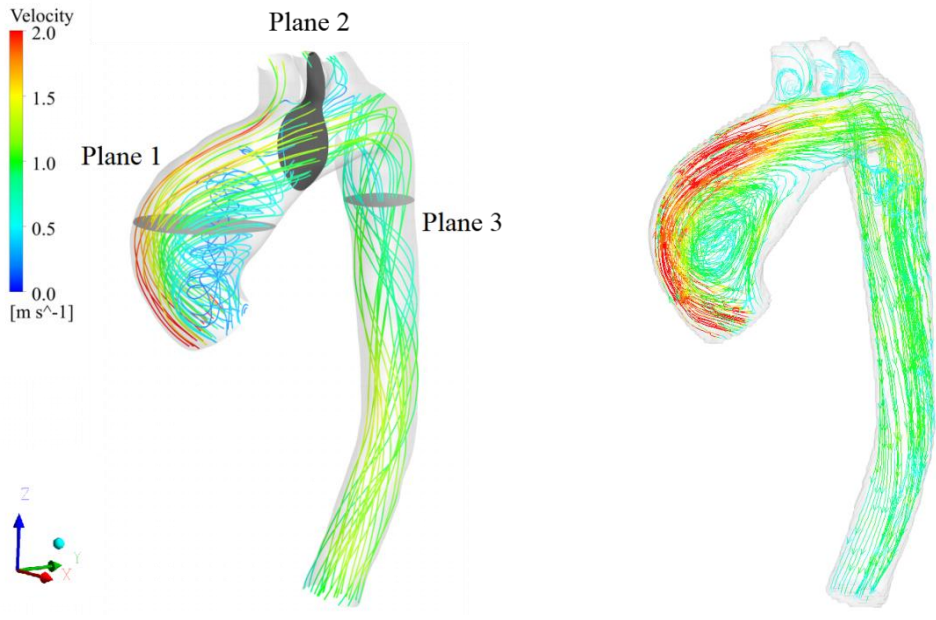

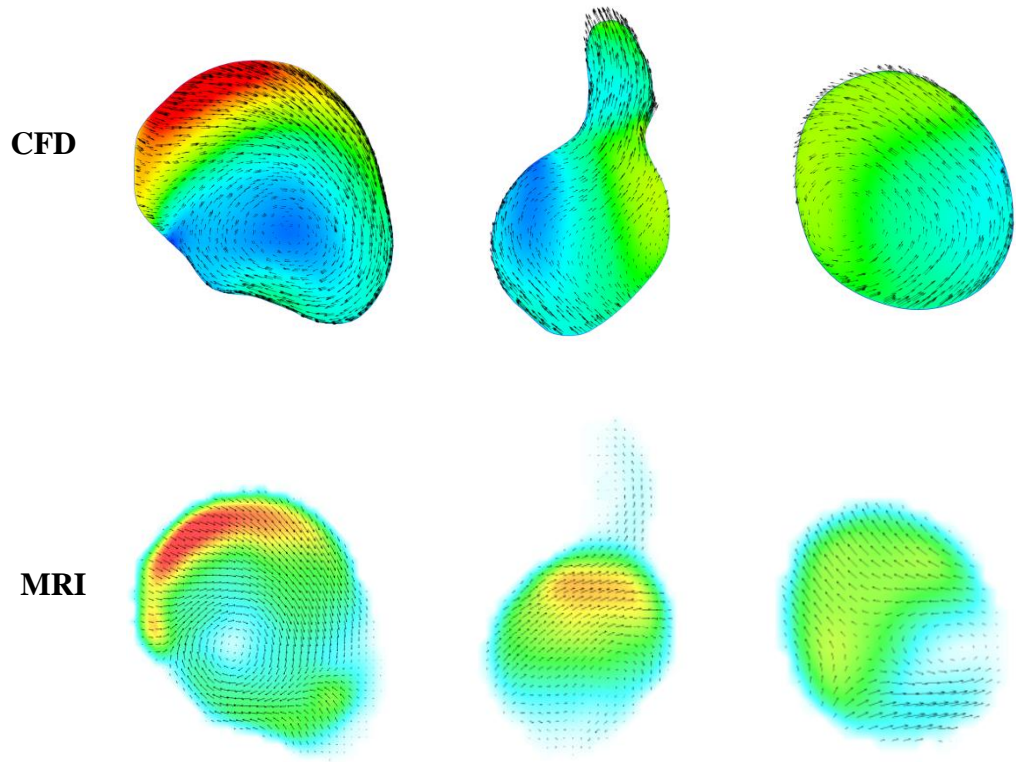

Fig. 6 Streamlines in aorta during the peak of systole. (a) 4D flow MRI processed by improved DFS, (b) CFD; and velocity vectors and contours in three planes for 4D flow MRI and CFD (c), respectively.

Table 2. Velocity correlation coefficient and average error in three planes between 4D flow MRI and CFD.

|             | Velocity u | Velocity v | Velocity w |
|-------------|------------|------------|------------|
| **Plane 1** | 0.8649     | 0.6469     | 0.9327     |
| **Error (m/s)** | 0.0192 | 0.0756     | 0.0278     |
| **Plane 2** | 0.5516     | 0.5690     | 0.4281     |
| **Error (m/s)** | 0.1224 | 0.2461     | 0.2935     |
| **Plane 3** | 0.7432     | 0.4317     | 0.8389     |
| **Error (m/s)** | 0.0957 | 0.0040     | 0.2185     |

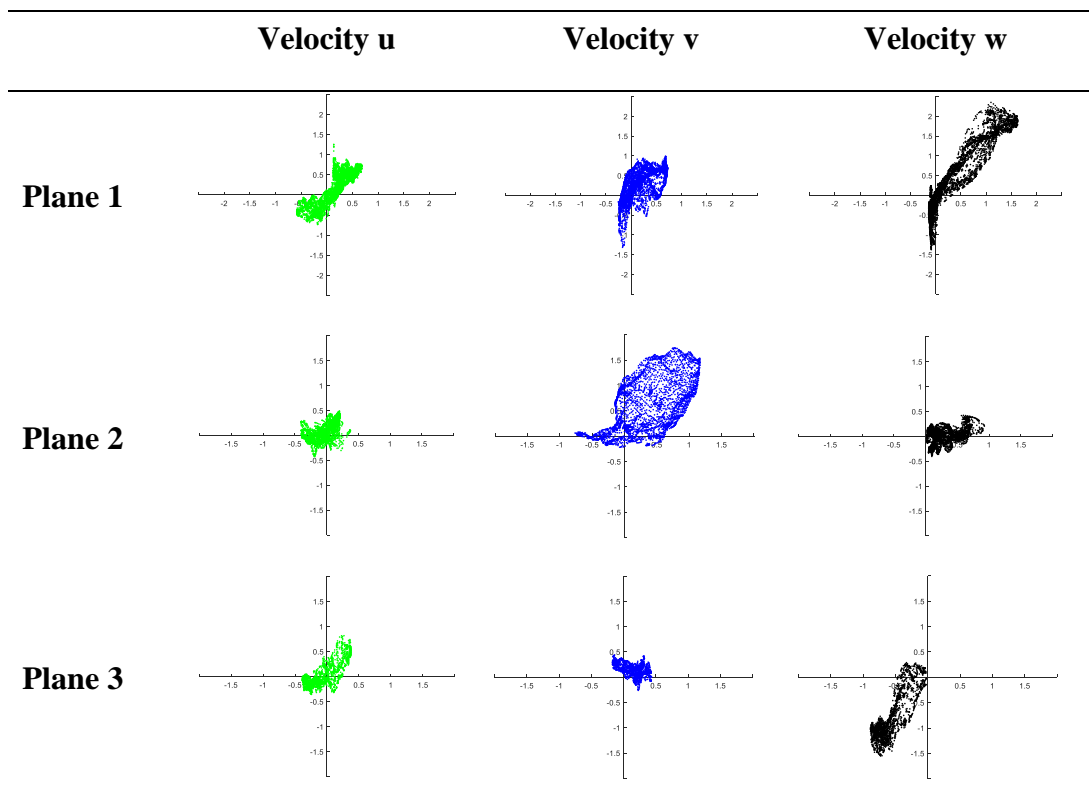

Fig. 7 The correlation distribution of velocity in three planes between 4D flow MRI and CFD. Horizontal axis represents CFD [m/s] and vertical axis represents MRI [m/s].

**Comparison of WSS with CFD simulation**

In order to further illustrate the advantages of the improved DFS method for velocity field optimization in aorta, WSS calculated by three different optimization methods and by CFD method are presented in Fig. 8. It can be seen that WSS obtained from the original 4D flow MRI is discontinuous (see Fig. 8(a, e)). Fig. 8(b, f) shows the WSS computed by smoothed velocity filed using traditional DFS method. The distribution of WSS gets smooth, but the magnitude of WSS in descending aorta becomes quite small and even distorted. It has a great improvement when improved DFS method is applied to ameliorate the velocity field, as shown in Fig. 8(c, g).

The WSS calculated by CFD is shown in Fig. 8(e, h). Generally, for WSS, there is a good consistency between 4D flow MRI with improved DFS method and CFD in most regions. Although WSS computed by 4D flow MRI is smaller than that by CFD at the entrance of ascending aorta, it reaches the maximum at the lateral side of ascending aorta for both of them. Gradually, the WSS decreases to the lowest value from ascending aorta to aortic arch. There are evident differences in supra-aortic vessels (BCA, LCCA, LSA) where WSS derived from CFD are much higher than that from 4D flow MRI. Instead, it shows a great similarity in descending aorta with the exception of the exit.

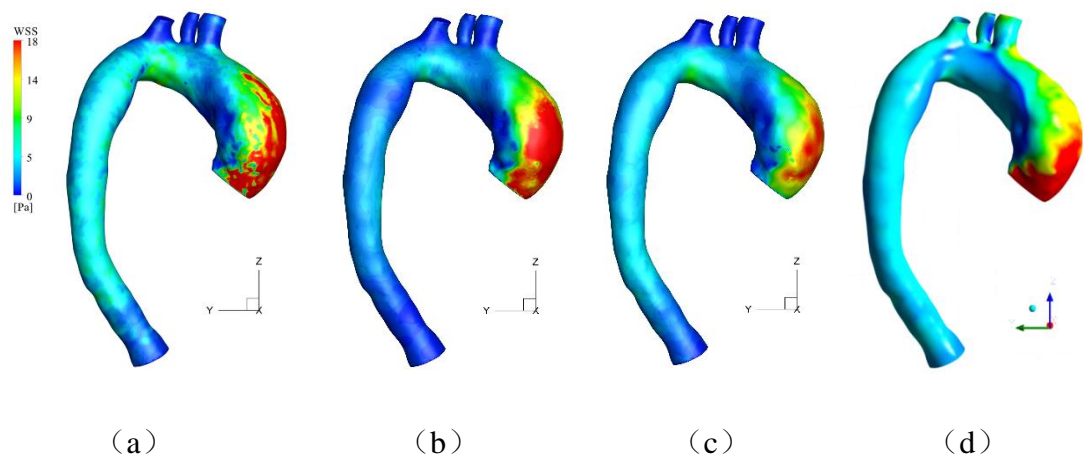

（a） （b） （c） （d）

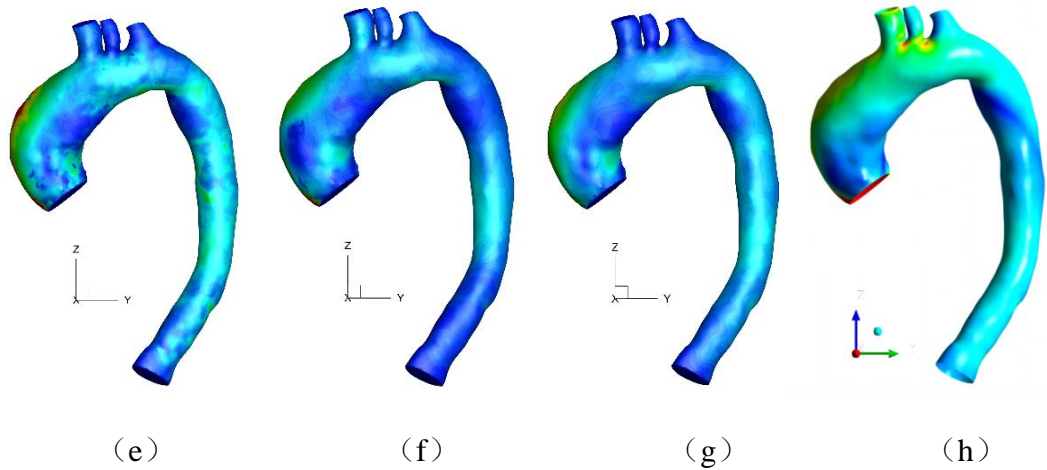

（e） （f） （g） （h）

Fig. 8 WSS comparison. (a, e) original 4D flow MRI, (b, f) 4D flow MRI optimized by traditional DFS method, (c, g) 4D flow MRI optimized by improved DFS method, (d, h) CFD.

## DISCUSSION

The original 4D flow MRI velocity field dramatically and vividly reveals the flow pattern in the mainstream area of aorta, but it contains lots of outliers in velocity field. The main reason is that the original 4D flow MRI velocity field contains plenty of noises, dead pixels and experimental errors. At the same time, velocity at wall can be observed, which was mainly caused by the low spatial resolution and partly by the dynamic motion of the aorta. 4D flow MRI velocity field smoothed by traditional DFS method shows good smoothness on velocity field and reduction on divergence, especially for the mean value of divergence. Nevertheless, velocity near wall is still uncorrected, and velocity magnitude in the core region of descending aorta apparently decreases in comparison with the original 4D flow MRI velocity field. To a great extent, the reason is that there is a shortage on numerical difference method when traditional DFS method is used to smooth and denoise the velocity field. That is to say, the velocity field optimized by traditional DFS method has an inferior fidelity in this case. 4D flow MRI velocity field smoothed by the improved DFS method has a great coincidence with the original velocity in the mainstream region, and it also improves the velocity profile near the wall. Velocity gradient is more obvious near the wall when no-slip wall boundary condition and an alternative optimization method in velocity flied are used under the ground of traditional DFS method. The advantage of improved DFS method can also be found from divergence value to the 4D flow MRI velocity field.

CFD has a strong advantage in calculating aortic hemodynamics, not only showing a high fidelity in comparison with in vivo and in vitro aortic velocity and pressure field[70,71], but presenting a good consistency in terms of hemodynamic parameters, such as WSS and turbulent kinetic energy as well[43,72]. Typically, blood flow is laminar in healthy arteries. However, turbulence aroused by high-frequency fluctuations can be found in vivo in ascending aorta, arteriovenous grafts and mechanical heart valves[73]. By comparing a few CFD results of different turbulent models with optimized 4D flow MRI data, it is found that the results calculated by k-epsilon RNG model show the highest correlation with the 4D flow MRI data[74]. These all illustrate the reliability of CFD calculation results and it can be used to verify the correctness and accuracy of 4D flow MRI data.

In the present study, there is a high-speed flow region with maximum velocity at the lateral side of ascending aorta and helical flow at the inner curvature for both data of 4D flow MRI and CFD during systole. The reason is that the formation of low-pressure zone and the emergence of flow separation at the ascending aorta near the inner side, which was mainly because the patient suffered from aortic regurgitation. In addition, the flow separation is accelerated by the twisted contraction of left ventricle. The high-speed flow region extends from the great curvature of the ascending aorta to the aortic arch, and the velocity gradually decreases due to the shunt of the supra-aortic trunks. Great similarity in velocity distribution derived from 4D flow MRI and CFD can be seen from plane 1. It can also be found from the high correlation in three velocity components, especially for the velocity along flow direction. The great similarity indicates that 4D flow MRI velocity field optimized by improved DFS method presents the intrinsic flow in aorta. At the same time, it demonstrates that the application of 4D Flow MRI velocity profile as boundary condition of CFD calculation completely restores the veritable flow pattern at the entrance of aorta during the peak of systole as well. On the contrary, the complex flow pattern caused by aortic valve lesion or left ventricular twisted contraction cannot be observed if mean mass flow or mean velocity value[15,75,76] is used as boundary condition for CFD calculations.

Basically, because of the exposure to the high-speed flow region in ascending aorta flow, the velocity is relatively large at the outer side of aortic arch. Meanwhile, the flow in the inner side of aortic arch shows slightly helical pattern due to the rotational bend of aortic arch in morphology[77]. However, the magnitude and the distribution of velocity have conspicuous discrepancy in aortic arch between 4D flow MRI and CFD. It can be inferred from the difference in plane 2. Velocity distribution in the aortic arch from 4D flow MRI is bilaterally symmetric, and velocity gradient from the outer side of aortic arch to LCCA is much higher. This is because the compliance of aorta in vivo leads to dilation of vessel wall during the systolic period. Therefore, the flow pattern in aortic arch region is caused partly by compliance and partly by the tortuous morphological structure in 4D flow MRI. Instead, the flow pattern from CFD in aortic arch is totally decided by the geometry when a fixed velocity profile is given. That is to say, CFD results may be not a good reference for assessing the reasonableness of 4D flow MRI velocity filed in the aortic arch in this study[74]. For the flow in supra-aortic vessels, Gallo et al. (2012) applied two different methods for the measurement of flow[45]: (a) flow rate was straightforwardly extracted at outlet section perpendicular to each vessel, and (b) flow rate was calculated by the difference between flow rates measured at two sections of the aortic arch, and the sections situated immediately in the upstream and downstream of each branch. It was found that the mass was not conserved, neither instantaneously nor as an average in strategy (a), and the flow rate measured by strategy (b) assured the instantaneous mass conservation constraint in the acquisition instant instead. Unsurprisingly, velocity magnitude was quite low and there was local inverse flow in supra-aortic arteries from 4D flow MRI in this paper because of the low resolution[38]. In other words, it proved that velocity filed indeed was underestimated to some extent in 4D flow MRI in supra-aortic vessels compared with CFD results. The correlation coefficient along z-axis in plane 2 was also illustrated that.

Flow pattern is comparatively monotonous in descending aorta and the velocity is approximately similar in both 4D flow MRI and CFD overall. As for streamlines in the upper segment of descending aorta in CFD, a twisted flow is shown because of the rigid wall boundary condition during simulation. In contrast, this phenomenon is overcome in 4D flow MRI due to the biological compliance. Nevertheless, subtle flow separation is discovered in the inner side of proximal segment. The correlation coefficient is pretty high with the exception of that along y-axis, indicating that velocity has a great agreement in the mainstream direction. In addition, The magnitude of velocity in the distal segment of aorta is lower in 4D flow MRI than that in CFD because of the mildly underestimated flow rate[74].

The magnitude and distribution of WSS is always one of the most concerned parts in research when aortic diseases are referred to. Various methods have been applied to dispose near-wall velocity profile in 4D flow MRI velocity field due to the problem of low spatial and temporal resolution[43,59,61,67], but WSS varies in a large extent and it is underestimated technically. An implicit wall function based on eddy viscosity model is used to optimize the velocity profile, and then WSS is calculated in the present study.

WSS calculated from the original MRI velocity field is discontinuous and patchy because of the existing of outlier. The velocity is manually set to be zero at the wall boundary during the calculation of WSS in original velocity field. Therefore, it causes the increase of velocity gradient in the near-wall region, which is responsible for the larger WSS. The results of WSS are ameliorated when traditional DFS method is used to optimize the velocity field partly. This is because noise is substantially reduced and velocity field satisfies the divergence-free condition at the same time. However, the traditional DFS method does not provide a near wall treatment satisfying boundary conditions, and the greatest difficulty in applying DFS to 4D flow MRI is how to deal with no-slip wall condition. To overcome the above-mentioned defect, an improved DFS method with wall boundary condition is introduced to amend the velocity field and gives the final WSS. It can be found that the distribution of WSS becomes smoother and the magnitude of WSS turns smaller correspondingly with reinterpolation of the interior velocity field.

It is well known that velocity extracted by 4D flow MRI is restricted by the spatial resolution of pixels and it would mislead the result of WSS. On the other hand, the spatial and temporal resolution of CFD is much higher than that in any vivo and vitro measurements. The reason is that temporal resolution is controlled by time steps in CFD and spatial resolution is defined by the computing mesh. This is the key to obtain more accurate hemodynamic index. For instance, WSS is partly relying on the spatial resolution of pixels in 4D flow MRI and the spatial resolution of mesh in CFD near wall region. Multiple investigations have been carried out in evaluating the helical flow and WSS with CFD methods in the cases of intracranial and cardiac diseases, such as bicuspid aortic valve and aortic stenosis[78], retrograde aortic type A dissection (RTAD)[79], and aortic lesions treated by stent-graft implantation[80]. It is found that high WSS has a strong correlation with the formation of aneurysms, and too low WSS can lead to rupture of aneurysms[81]. By analyzing and judging the blood flow pattern, it showed that WSS has a great correlation with the morphology of the aneurysm[82].

At the peak of systole, except for the area close to the boundaries, the magnitudes of WSS calculated by 4D flow MRI and by CFD are basically the same and the distribution of WSS also shows high similarity. This demonstrates that the improved DFS method can greatly optimize the velocity profile near the wall and enhance the accuracy of WSS. It should be noted that the velocity profile imposed on entrance of CFD computing model as inlet boundary condition is not optimized, and this is part of the reason for the higher WSS at inlet surface in CFD. In addition, the inelastic wall boundary condition in CFD may partly overestimate the WSS because a wall with biological compliance is known to reduce WSS[68,83,84]. WSS reaches the maximum at the outer side of ascending aorta in both 4D flow MRI and CFD because utmost velocity gradient presents. Due to the helical flow, relatively small WSS is shown at the lesser curvature of aorta. WSS gradually decreases from ascending aorta to the aortic arch, where it is symmetrical in a way in 4D flow MRI. On the contrary, WSS is larger in the interior side than that in the posterior side in CFD. The different patterns of WSS from 4D flow MRI and CFD present various flow separation in the ascending aorta. It also illustrates the importance of vessel compliance when blood flow is simulated in big arteries by CFD. Although WSS calculated by 4D flow MRI has been greatly improved, it is still underestimated at supra-aortic arteries because of the underestimated velocity field. It means that velocity profile is still limited by the insufficient resolution and it is difficult to accurately compute hemodynamic parameters at supra-aortic trunks with 4D flow MRI. In contrast, CFD presented a relatively high WSS. But it does not mean that WSS calculated by CFD is totally correct because WSS may be overestimated in CFD as previously mentioned. Low WSS area at the proximal of descending aorta can be found in 4D Flow MRI because of the flow separation caused by compliance. This is also discovered in CFD due to the presence of swirling flow resulted from the inflexible and slightly twisted morphology of aorta. The distribution of WSS shows the same trend like supra-aortic trunks. In summary, the mismatching of WSS between 4D flow MRI and CFD at boundary area is partly caused by underestimated velocity filed in 4D flow MRI and partly by the overestimation of inelastic wall boundary condition in CFD. Apart from this, there is a good consistency between 4D flow MRI and CFD. It demonstrates that velocity distribution is restored well and aortic WSS is correctly calculated to a large extent based on 4D Flow MRI, which is helpful for judging the location and the possible development of lesions in diseases.

# CONCLUSIONS

Two different methods, traditional DFS and DFS with wall boundary condition, are applied to optimize the original velocity field, and velocity field computed by CFD is used as a reference at the same time. It is found that the improved DFS can correctly display the near-wall velocity profile in most regions. Meanwhile, improved DFS can greatly reduce the divergence of the entire flow field, so that flow field satisfies the no-slip condition on the wall and divergence-free requirement. By the comparison of velocity field in 4D flow MRI and CFD, strong correlation is found to exist in most regions except for the velocity distribution in aortic arch and supra-aortic arteries. Compliance in vivo leads to the apparent difference between 4D flow MRI and CFD in velocity distribution in aortic arch. For the flow in supra-aortic vessels, velocity calculated by 4D flow MRI is always underestimated due to the limitation of resolution. WSS calculated from the original 4D flow MRI velocity field is discontinuous because of the existing of outlier and noise in the flow field. Considering the ability of near-wall treatment, improved DFS method is superior to traditional DFS method in the computing of WSS. It uses an implicit wall function with higher precision to calculate aortic WSS. The results show a relatively high similarity in WSS computed by CFD. Inevitably, there are certain differences of WSS between 4D flow MRI and CFD in aortic boundaries and supra-aortic vessels. This is partly caused by insufficient revolution in several boundaries in 4D flow MRI and partly by the overestimation of inelastic wall boundary condition in CFD.

There are several limitations in the current work. First, this study is carried out considering only one image-based diseased aorta. More healthy or diseased subjects should be enrolled for illustrating the advantages of improved DFS method and the reliability of near-wall function. But for the optimization of velocity field and WSS with improved DFS method in 4D flow MRI, the strategy shows promising results compared with CFD simulation. Second, blood flow is simulated only at the peak of systole. It is not possible to simulate the internal flow characteristics of the aorta in one cardiac cycle at this circumstance. If a flat velocity profile changing over cardiac cycle is imposed on the inlet, it does not perform the high-speed flow region and helical flow aroused from the physiological diseases at all. Accordingly, there is no much point even if elastic aorta wall can be properly simulated using a uniform value in CFD in this study. The reconstruction of aorta is on the basis of grayscale information and velocity field, so it fails to build the aortic computing model during the diastole because of the low velocity field at this period. Therefore, it is meaningful when simulation is going at the peak of systole and may be at some phases close to the peak. Third, average resistance obtained from some previous investigations is imposed on the outlet boundaries of computing model. As is known, CFD results are severely relying on the boundary condition, so patient-specific boundary conditions are vital to get the desirable simulation results. But in order to reduce the economic cost, average resistance is introduced and this still acquires well-matched outcomes compared with 4D flow MRI.